\documentclass[showpacs,amsfonts,twocolumn,aps,prl]{revtex4} 

\newcommand\NPA{{Nucl. Phys.} A}
\newcommand\NPB{{Nucl. Phys.} B}

\newcommand\PLB{{Phys. Lett.} B}

\newcommand\PRL{Phys. Rev. Lett.}

\newcommand\PRD{{Phys. Rev.} D}

\newcommand\IJMPA{{Int. J. Mod. Phys.} A}

\newcommand\CQG{Class. Quant. Grav.}

\newfam\BMath
\font\BMathL=cmmib10 
\font\BMathl=cmmib7
\font\BMathm=cmmib5
\textfont\BMath=\BMathL \scriptfont\BMath=\BMathl
\scriptscriptfont\BMath=\BMathm

\renewcommand\a{\alpha}

\renewcommand\d{\delta}
\newcommand\e{\epsilon}

\renewcommand\l{\lambda}
\newcommand\m{\mu}
\newcommand\n{\nu}

\newcommand\p{\pi}
\newcommand\s{\sigma}
\renewcommand\t{\tau}

\newcommand\f{\phi}

\renewcommand\j{\psi}
\renewcommand\o{\omega}

\newcommand\cb{{\cal B}}

\newcommand\cf{{\cal F}}

\newcommand\ch{{\cal H}}

\renewcommand\exp{\mbox{\rm exp}}  
\newcommand\tr{\mbox{\rm tr}} 

\newcommand\ra{\rightarrow}

\newcommand{\lan}{\langle}     
\newcommand{\ran}{\rangle}     

\newcommand\del{\partial}

\newcommand{\half}{\frac{1}{2}}

\newcommand\be{\begin{equation}}
\newcommand\ee{\end{equation}}
\newcommand\bea{\begin{eqnarray}}
\newcommand\eea{\end{eqnarray}}
\newcommand\beal{\begin{align}}
\newcommand\eeal{\end{align}}

\newcommand\ba{\begin{array}}
\newcommand\ea{\end{array}}
\newcommand\bc{\begin{center}}
\newcommand\ec{\end{center}}

\newcommand\eref[1]{Eq.~(\ref{#1})}

\newcommand\bfi{\begin{figure}}
\newcommand\efi{\end{figure}}
\newcommand\bpi[1]{\begin{picture}#1}
\newcommand\epi{\end{picture}}

\def\jou#1#2#3#4{{#1} {\bf #2}, #3 (#4)}

\def\bem{\begin{pmatrix}}
\def\eem{\end{pmatrix}} 

\newcommand\ve{\varepsilon}

\begin{document}

\title{What exactly is a Skyrmion?} 

\author{S.M.H. Wong} 
\affiliation{Department of Physics, The Ohio State University, Columbus, 
Ohio 43210 
}

\begin{abstract} 
Skyrmions are well known to be baryons because their topological charge 
has been positively identified with the baryon number. Beyond that their 
identity has never been clear. In view of the possibility of skyrmion 
production through Disoriented Chiral Condensates in heavy ion collisions, 
the exact identity of the skyrmion must be resolved before they can be 
identified in experiments. It is shown that skyrmions are not individual 
baryons but coherent states of known baryons and higher resonances on a 
compact manifold associated with the spin and flavor symmetry group. 
An outline of how to calculate exactly the probability amplitudes of the 
superposition of physical baryon and excited baryon states that make up 
the skyrmion is given. 
\end{abstract} 

\date{24 February 2002} 



\maketitle

\section{Introduction}
\label{s:intro}

Over a decade ago it was suggested that in heavy ion collisions
classical pion field may be formed, this would lead to coherent 
pion production \cite{aa}. Motivated by the Centauro events 
large fluctuation in the ratio of the number of charged-to-neutral pions 
was predicted to occur from event to event in heavy ion collisions 
\cite{aa,bj1,bk}. More careful investigation showed that large 
domains with different chiral orientations were necessary for 
any observably large fluctuations in the pion ratios to occur 
\cite{rw1}. Such a phenomenon was coined Disoriented Chiral Condensates 
(DCC) \cite{rw2,bkt}. A number of papers were devoted to looking  
for DCC at the Tevatron \cite{bkt,mmx} and a even greater number were 
devoted to finding them in heavy ion collisions. Unfortunately so far all 
searches for them ended in failure at the Super Proton Synchrotron (SPS)
at CERN \cite{wa98} and at the Fermilab Tevatron from the MiniMax 
experiment \cite{mmx}. Facing these results it is all too easy to conclude that 
there is no DCC formation at all in these experiments. Nevertheless one has 
to bear in mind that in heavy ion collisions pions are the most copiously 
produced hadrons. In order to observe DCC a large number of coherent pions 
must be produced so that they can rise above this background. Large domains 
with fixed chiral orientation are required for this to happen \cite{rw1,gm}. 

Recently the possibility of baryon-antibaryon productions through small
domain DCC was raised and connected to the $\Omega$ and $\bar \Omega$
data from the SPS \cite{kw}. This occurs through skyrmion  
and antiskyrmion formation \cite{sk}. This possibility in the context of 
heavy ion collisions was previously raised in \cite{dg,ehk,ks}. However 
it was first connected to DCC and data only recently in \cite{kw}. In it a 
problem on how to confront experimental data was encountered. Although 
a skyrmion is generally known to be a baryon or nucleon, its exact identity 
has never been clear. However for serious phenomenological applications such 
inexactitude cannot be tolerated. In this paper the mystery to the exact 
identity of the skyrmion is revealed. They are coherent states of baryons 
and excited baryons. An outline will be given on how to connect and obtain 
the physical states for a skyrmion. The details will be presented elsewhere
\cite{me}.

\section{The Skyrme Model}
\label{s:sk-md}

In 1962 Skyrme introduced the following lagrangian 
\be {\cal L}_S = \frac{f_\p^2}{4} \; \tr(\del_\m U\del^\m U^\dagger) 
           +\frac{1}{32g^2}  \; \tr[U^\dagger \del_\m U,U^\dagger \del_\n U]^2
\label{eq:l_s}
\ee 
where 
\be U = \exp \{i \mbox{\boldmath $\t$} \cdot \f /f_\p\} 
      = ( \s + i \mbox{\boldmath $\t$} \cdot \mbox{\boldmath $\p$})/f_\p 
        \; ,
\ee
$f_\p$ is the pion decay constant and $g$ is now known to be the
$\rho$-$\p$-$\p$ coupling. The first term is the usual non-linear sigma
model and the second part was introduced by Skyrme \cite{sk}. 
This model belongs to a family of lagrangians that are known to approximate 
QCD at low energies. Skyrme found a family of classical static solutions to the 
equation of motion derived from ${\cal L}_S$. They can be written in the form  
\be U = U_S = \exp \{i \mbox{\boldmath $\t$}\cdot \mbox{\boldmath $\hat r$}\; 
                     F(r) \}
\label{eq:u_s}
\ee
where $F(r)$ is a radial function which must satisfy certain specific 
boundary conditions. They are 
\be F(r \ra \infty) \ra 0  \mbox{\hspace{.8cm} and \hspace{.8cm}} 
    F(r=0) = N \p \; 
\ee 
\cite{sk,bha,bmss}. $N$ in the last expression is the integral valued
topological charge or winding number. It has been identified as the baryon 
number \cite{sk,wi} provided that the Wess-Zumino effective action  
\cite{wz} is included to eliminate the unphysical symmetry inherent in 
the non-linear sigma model. Only $N=1$ will be considered in this paper 
which corresponds to the case of one baryon. This connection of 
skyrmion to baryon has generated much applications: from being the model 
for studying baryon mass spectrum \cite{anw,gu,wk} to that for studying 
the quark, spin content and form factors of the nucleon \cite{med}, from 
being a non-linear mechanism for baryon production in jets \cite{ek} to 
that in heavy ion collisions \cite{dg,ehk}. There are many more works for 
generalizing to more than two quark flavors. A nice review of many of these 
applications and generalization can be found for example in \cite{hw}
and in references therein. 

Although the identification of a skyrmion as a baryon was made early
\cite{sk} and confirmed in \cite{wi,bnrs}, there was no direct link of 
a skyrmion to a nucleon. It was speculated in \cite{wi} that the ground 
state of the skyrmion is the nucleon. Indeed apart from the baryon
number, a skyrmion has no quantum numbers whereas baryons have spin 
and isospin. To obtain these quantum numbers, a skyrmion has to be
quantized.

\section{Quantization of the $N=1$ Skyrme Hamiltonian} 
\label{s:qt}

In this paper only flavor $SU(2)$ skyrmion will be considered. 
Quantization of skyrmion was done in \cite{anw} using the techniques 
of collective coordinate. This amounts to finding time-evolution
about the static solution. Substituting the solution \eref{eq:u_s} for
$N=1$ into the lagrangian density \eref{eq:l_s} and integrating over space 
gives the energy or mass of the static skyrmion $M=-\int d^3 r {\cal L}_S$. 
The collective coordinate is introduced by observing that if $U_S$ is a 
solution of the Euler-Lagrange equation to \eref{eq:l_s} so is 
$U_S'=A(t)U_S A^\dagger(t)$. Here $A(t)$ is an element of the group $SU(2)$, 
which may be more conveniently written as
$A=a_0 + i \text{\boldmath $a$} \cdot \text{\boldmath $\t$}$. 
The $a_i$'s, $i=0,1,2,3$. satisfies the constraint $a_0^2 +${\boldmath $a$}$^2=1$ 
as for any element of $SU(2)$. Substituting now the new $U_S'$ into \eref{eq:l_s} 
and integrating over space gives the lagrangian in terms of the $a_i$'s as 
\be L = 2\l \sum_{i=0}^3 (\dot a_i)^2 - M  \;. 
\label{eq:sL2} 
\ee 
$\l$ and $M$ are two constants and their explicit forms are unimportant 
here. In this form, the conjugate momentum to $a_i$ is readily 
found to be $\p_i = 4 \l \dot a_i$. A naive quantization procedure 
$[a_i,\p_j]=i\d_{ij}$ permits the equally naive representation 
$\p_i = -i \del_i$ \cite{fn1,di}. $\del_i$ is short for $\del/\del a_i$. 
It works nevertheless if one is careful with the differentiation. 
The Hamiltonian becomes 
\be H = \frac{1}{8\l} \sum_{i=0}^3 \p_i^2   +M 
      = -\frac{1}{8\l} \sum_{i=0}^3 \del_i^2 +M  
\label{eq:ham}
\ee
where the $\nabla^2 =\sum_i \del_i^2$ is understood to be acting on 
variables constrained on $S^3$. The wavefunctions of this Hamiltonian are 
in general polynomials of $(a_i+i a_j)$. For example with integral power $l$,  
$-\nabla^2 (a_i+i a_j)^l = l(l+2) (a_i+ia_j)^l$ 
\cite{anw}. One could also write $\nabla^2$ in terms of the spin
and isospin operators 
$-\nabla^2 = 2 \sum_{i=1}^3 (J_i^2+I_i^2)$. 
These are for $i=1,2,3$ 
\bea I_i &=& \half i \Big (a_0\del_i -a_i\del_0 -\ve_{ijk} a_j \del_k \Big ) \\
     J_i &=& \half i \Big (a_i\del_0 -a_0\del_i -\ve_{ijk} a_j \del_k \Big ) \;.
\eea
The Hamiltonian commutes with any of these 
$[H,J_i] = [H,I_i]= [I_i,J_k]=0$, 
so simultaneous eigenstates of energy, spin and isospin can be found. 
Favoring the third direction as usual 
\be J_3 (a_1 \pm i a_2)^l = I_3 (a_1 \pm i a_2)^l 
                          =\pm \mbox{$\frac{l}{2}$} (a_1 \pm ia_2)^l  
\ee
so odd integer $l$ gives half-integral spin and isospin states while
even integer $l$ gives integral spin and isospin states. This system
permits both bosonic and fermionic states. Since baryons are fermions 
so one can have only polynomials of odd degree $l$ for physical states. 
For example labeling states by their total $j$, spin $m$ and isospin $n$
quantum numbers $|jmn\ran$, the simplest proton and neutron spin up 
wavefunctions are polynomials of degree 1  
\bea \lan a |p \uparrow   \ran &=& \lan a|\mbox{$1/2,1/2,1/2$} \ran 
                                =  \frac{1}{\p} (a_1+i a_2)  \\
     \lan a |n \uparrow   \ran &=& \lan a|\mbox{$1/2,1/2,-1/2$} \ran 
                                =  \frac{i}{\p} (a_0+i a_3)  \;. 
\eea 
where the Dirac ket notation $|a\ran$ is used to represent the position
vector on $S^3$. In general the wavefunctions are holomorphic 
functions on $S^3$ in terms of the coordinate $a_i$. One could easily 
write down the wavefunctions for the $\Delta$'s and other excited baryon 
states. With the collective coordinate, the quantization of this system 
is quite straightforward. 

Although the physical states have been produced by quantization and they 
possess the required quantum numbers, this does not furnish a connection to 
the classical skyrmion. Quantization by itself does not connect the 
quantum mechanical to the classical but if one recalls the problem of
the simple harmonic oscillator, Schr\"odinger showed a long time ago that 
classical-like solutions could be found among the quantum states. 
These are the so-called coherent states. They are the closest quantum
equivalence to the classical solutions. Our task is now reduced to
finding the coherent states which are superposition of the physical 
quantum states.

\section{The Criteria for Coherent States} 
\label{s:cs} 

Very unlike the n-dimensional harmonic oscillator where the space is
$R^n$, our problem resides on the surface of the four-dimensional
sphere which is a compact manifold. The techniques used in the harmonic 
oscillator problem cannot be applied straightforwardly. One should however
be able to draw close analogy and use that as a guide. There are many
works in the literature on coherent states, in curved space-time, compact
manifold with various applications such as in quantum gravity, see 
for example \cite{ap,c-pro,tt,hm} and references therein. There are
a variety of ways to construct them and they are not all equivalent. 
It is therefore best to choose the method that is closest to the original
and is an analog of that in compact spaces. In our case our space is 
$S^3$ and we would like to construct coherent states on it.
The case of coherent states on a circle and a sphere have been done
in \cite{kr} and generalized to $S^n$ in \cite{hm} using the techniques 
of the generalized Segal-Bargmann transform or ``coherent state''
transform \cite{tt,hm}. 

According to \cite{tt,hm} coherent states are defined as simultaneous 
eigenstates of some annihilation operators as in the usual harmonic 
oscillator problem. From quantum mechanics, we have 
\be A_i = \mbox{$\frac{1}{\sqrt{2m\o\hbar}}$} (m\o X_i +i P_i) 
\label{eq:ai-sho}
\ee
with $A_i |\a \ran = \a_i |\a \ran$. 
They are labeled by points in phase space. Again the same is true in the 
original because the usual coherent states are labeled by the set of
eigenvalues $\a_i$'s. These are linked to the expectation values
\be \lan X_i \ran =   \sqrt{\mbox{$\frac{\hbar}{2m\o}$}}  (\a_i +\a_i^*) \;\;, 
    \lan P_i \ran =-i \sqrt{\mbox{$\frac{\hbar m\o}{2}$}} (\a_i -\a_i^*) 
\ee
which is equivalent to labeling by phase space points. In Euclidean space 
$R^n$ the method of constructing the annihilation operators should reduce to 
the well known form. This has been shown to be true in \cite{hm}. Finally 
the equivalence of $\lan \Delta X_i^2\ran$ and $\lan \Delta P_i^2\ran$ should be 
constants independent of quantum numbers as in the simple harmonic oscillator. 
This was considered in \cite{kr} and they largely hold true. For these 
reasons it is most reasonable to apply their method for constructing the 
annihilation operators and the coherent states. Note that the construction
advocated in \cite{ap} does not fulfill some of these criteria and the
resulting coherent states are not direct generalization of those from the
well known ones.

\section{Problem in the Straight forward Coherent State Construction}
\label{s:p} 

The construction of coherent states can be summarized as follows. 
The method of \cite{tt,hm} gives the annihilation operators on a compact
manifold as 
\be A_i = e^{-C} X_i e^C   
\label{eq:ai}
\ee
where $X_i$'s are the $i$ component of the coordinate operator on the compact 
manifold and $C$ is a dimensionless complexifier. The latter is so called
because \eref{eq:ai} is a transformation of real phase space coordinates
($X_i$,$P_i$) into the complex pair ($X_i^{\mathbb C}$,$P_i^{\mathbb C}$). 
$A_i=X_i^{\mathbb C}$ is the complexified $X_i$ and $C$ can be thought of
as the generator of the transformation. On $S^n$ the $A_i$ operators 
can be rewritten in the form 
\be A_i = \s_x(J) X_i +i \s_p(J) P_i     
\ee
in analogy to \eref{eq:ai-sho}. The angular momentum or spin $J$ dependent 
coefficients $\s(J)$ are a feature of the compact $S^n$ \cite{hm,kr}. 
Equipped with the annihilation operators, simultaneous eigenstates of them 
can be constructed from a fixed position state vector $|x\ran$ on the manifold. 
By definition 
$X_i |x\ran = x_i |x\ran$, 
then if we let $|\j \ran = e^{-C} |x\ran$, 
this must be a simultaneous eigenstate of $A_i$ because 
$A_i |\j \ran = x_i |\j \ran$. 
The coherent state construction is almost complete. Because $x_i$ is 
real, one must now analytically continued $|x\ran$ and $x_i$ to complex 
coordinates $|x^{\mathbb C}\ran$ and $x^{\mathbb C}$. Once this last step 
is done, $|\j \ran$ is now the coherent state labeled by $x^{\mathbb C}$.
Therefore there is one coherent state per phase space point as in 
the usual harmonic oscillator. 

The actual application of this method to our problem is not so simple. 
Remember that on $S^3$ the coordinates are the $a_i$'s and we are interested
in coherent states expanded in terms of physical states. This can be
done by inserting a complete set of states, both integral and half
integral quantum numbers ($j,m,n$), in the above $|\j\ran$ 
\be |\j \ran = e^{-C} |a^{\mathbb C}\ran 
             = e^{-C} \sum_{j,m,n} |j,m,n\ran \lan j,m,n |a^{\mathbb C}\ran  
    \;. 
\label{eq:cs}
\ee
$\lan j,m,n |a^{\mathbb C}\ran$ is the complex conjugate of the 
wavefunctions evaluated at the complex coordinate $a^{\mathbb C}$ and
$C$ is related to the kinetic term of the Hamiltonian \eref{eq:ham} \cite{me}. 
This is highly undesirable because a skyrmion should not involve unphysical 
states. The simplest solution is to discard these from \eref{eq:cs}. 
Unfortunately this must fail. To see this, let us momentarily use a 
representation that states are wavefunctions and operators act by multiplication 
and/or differentiation. Then with only physical states, we have 
\be \j = e^{-C} \sum_{j,m,n \in \mathbb Z +\text{$\half$}} \j_{jmn} \;.
\ee
Acting on this with the position operator $\hat a_i$ gives
$\hat a_i \j = a_i \j$. 
Remember that physical state wavefunctions are odd degree polynomials. 
The above operator turns all odd into even polynomials. $|\j\ran$ 
without the unphysical states is no longer eigenstates of $A_i$ and 
therefore not a coherent state. So straightforward application of the 
method to our physical problem fails immediately.

\section{A Skyrmion is a Coherent State of Baryons and Excited Baryons}
\label{s:cs-b}

If we write the complete Hilbert space as a sum of fermionic and
bosonic space $\ch = \cf+\cb$, then the problem with the appearance 
of non-physical states is directly linked to the fact that $\hat a_i$ 
maps $\cf$ to $\cb$ and $\cb$ to $\cf$: 
$\hat a_i |\cf\ran = |\cb\ran$, $\hat a_i |\cb\ran = |\cf\ran$. 
If only one could use the operator products $\hat a_i^2$ or $\hat a_i \hat a_j$ 
as coordinate operators instead then the problem would disappear. These 
map $\cf$ to itself and likewise for $\cb$. This should allow us to 
discard $\cb$. How can one justify using such product operators? In any case 
instead of four $\hat a_i$'s, there would be nine independent combinations
of $\hat a_i \hat a_j$. The solution to this is instead of using the coordinate 
on $SU(2)$, one should use instead those of $SO(3)$. In other words, one first 
use the map of $SU(2)$ to $SO(3)$. An element of the former $A$ can be mapped 
to one of the latter using   
\be  A \t_i A^\dag = \t_j R_{ji}(A)   
\label{eq:su22so3}
\ee
\cite{bnrs}. Here $R$ is the $3\times 3$ rotation matrix. This
map gives 
\be R_{ij} = 2 a_i a_j +\d_{ij} (2 a_0^2-1) -2 \e_{ijk} a_0 a_k 
\ee 
where $i,j=1,2,3$ \cite{me}. The new coordinate with nine components 
represents points on the $SO(3)$ manifold and is now combinations of 
products of $a_i a_j$ in disguise. 

One will have to quantize the Hamiltonian again in terms of $R$ 
instead of the $a_i$'s \cite{me}. With these new coordinates, one 
can now follow the above prescription to construct the 
coherent state involving only physical baryon and excited baryon states. 
The relative probability of the physical states depends on the
operator $e^{-C}$ which acts on the physical states and the 
wavefunction evaluated at a complex point $R^{\mathbb C}$. From
the above discussion, these can be calculated. The make up of a skyrmion
as well as the relative probability a physical baryon state will be  
produced from skyrmion formation through DCC in heavy ion experiments 
can now be determined \cite{me}. 

The solution to the identity of the skyrmions naturally generates 
a lot of questions in light of the many existing literature and 
applications. The detail construction of a skyrmion in terms of baryonic 
coherent state, their application to heavy ion collisions, as well as other 
related questions will be pursued elsewhere \cite{me}. This completes the 
outline of how to arrive at a solution to the identity of a skyrmion. 
In brief a skyrmion is not a single baryon, but a quantum mechanical 
superposition of baryon and resonance states.

\section*{Acknowledgments}

The author thanks K. Kowalski and J. Rembieli\'nski for pointing out
ref. \cite{hm}, B.C. Hall for clarifying the annihilation operators, 
P. Ellis, U. Heinz and J. Kapusta for comments. This work was 
supported by the U.S. Department of Energy under grant no. DE-FG02-01ER41190.


\begin{thebibliography}{99}
\bibitem{fn1}One should really quantize the Dirac bracket instead of 
the Poisson bracket in the presence of constraints \cite{di}. This 
complication will be ignored in this paper.
\end{thebibliography}
\end{document}